\documentclass[a4paper]{article}

\usepackage{INTERSPEECH2021}
\usepackage{url, multirow}

\title{Joint Online Multichannel Acoustic Echo Cancellation, Speech Dereverberation and Source Separation}
\name{Yueyue Na, Ziteng Wang, Zhang Liu, Biao Tian, Qiang Fu}
\address{Alibaba Group}
\email{yueyue.nyy, ziteng.wzt@alibaba-inc.com}

\begin{document}

\maketitle
\begin{abstract}  
	This paper presents a joint source separation algorithm that simultaneously reduces acoustic echo, reverberation and interfering sources. 
	Target speeches are separated from the mixture by maximizing independence with respect to the other sources.
	It is shown that the separation process can be decomposed into cascading sub-processes that separately relate to acoustic echo cancellation, speech dereverberation and source separation, all of which are solved using the auxiliary function based independent component/vector analysis techniques, and their solving orders are exchangeable.
	The cascaded solution not only leads to lower computational complexity but also better separation performance than the vanilla joint algorithm.
\end{abstract}

\section{Introduction}

Smart devices that work in full-duplex speech interaction mode need to handle playback echos, room reverberation and interfering sources simultaneously. The three types of distortions are widely investigated in the literature and many classical algorithm have been developed separately, such as the normalized least mean square (NLMS) algorithm~\cite{shynk1992frequency,valin2007adjusting} for acoustic echo cancellation (AEC), the weighted prediction error (WPE) algorithm~\cite{nakatani2010speech,yoshioka2012generalization} for speech dereverberation (DR) and the auxiliary-function based independent component/vector analysis (Aux-ICA/IVA) algorithm~\cite{ono2010auxiliary,ono2011stable} for blind source separation (BSS). Joint solutions that consider two or three types of distortions are appealing, especially for real world applications, and could bring performance improvements over separate algorithms~\cite{takeda2009ica,takeda2012efficient,yoshioka2010blind,kagami2018joint,boeddeker2019jointly,nakatani2020computationally,nakatani2020jointly,ikeshita2021independent,cohen2018joint,togami2014simultaneous,carbajal2020joint,liu2021neural}.

Takeda et al.~\cite{takeda2009ica,takeda2012efficient} achieve both blind dereverberation and echo cancellation by applying a separation model of frequency domain ICA, which uses observed signal independence that holds under multiple input/output inverse filtering theorem (MINT) conditions. The authors also develop techniques to reduce the computational cost for their barge-in-able robot application.

Yoshioka et al.~\cite{yoshioka2010blind} propose a conditional separation and dereverberation (CSD) method, in which the separation and prediction matrices are alternately optimized with each depending on the other.
Boeddeker et al.~\cite{boeddeker2019jointly} propose a weighted power minimization distortionless response (WPD) beamformer that can perform simultaneous denoising and dereverberation. The beamformer is optimized under a single likelihood maximization criterion and shows superiority over the conventional cascade of WPE and a minimum-power distortionless response (MPDR) beamformer.

Several joint approaches have been proposed to perform AEC, DR and BSS at the same time~\cite{togami2014simultaneous,carbajal2020joint}. Togami and Kawaguchi~\cite{togami2014simultaneous} combine speech dereverberation, noise reduction and acoustic echo reduction in a unified framework by assuming a time-varying local Gaussian model of the microphone input signal. The algorithm parameters are iteratively optimized based on the expectation-maximization approach to calculate a minimum mean squared error estimate of a desired signal. Carbajal et al.~\cite{carbajal2020joint} further introduce a neural network to model the short term spectra of the target and residual signals after echo cancellation and 
dereverberation. 

Our previous work~\cite{wang2020semi,wang2021weighted} revisits the problem of DR and AEC, respectively, and proposes Aux-ICA based source separation approaches to each. This paper further proposes to jointly perform AEC, DR and BSS from a unified source separation perspective, assuming mutual independence of the mixing sources.
A joint source separation algorithm is first presented, which however comes at a high computational cost. 
We then decompose the separation matrix heuristically, and divides the joint optimization problem into sub-problems that can be tackled sequentially. 
The sequential cascaded solution not only leads to lower computational complexity but also better separation performance, due to relaxation of the assumptions made in the joint algorithm.

The rest of this paper is organized as follows. In Section 2, we formulate the problem using a convolutive signal model. The joint source separation algorithm and cascaded solutions are presented in Section 3. Experiments and concluding remarks are respectively given in Section 4 and Section 5.

\section{Problem formulation}

We consider a multi-channel convolutive mixture in the short-time Fourier transform (STFT) domain. An array of $M$ sensors captures signals from $N$ near-end sources ${\bf s} = [s_1,s_2,...,s_N]^T$ and $R$ far-end sources ${\bf r}=[r_1,r_2,...,r_R]^T$, where $(\cdot)^T$ denotes transpose. The sensor signals ${\bf x}=[x_1,x_2,...,x_M]^T$ are given by:
\begin{equation}\label{eq:signalmodel}
  {\bf x}(t,f) = \sum_{l=0}^{\infty}{\bf A}_l {\bf s}(t-l,f) + \sum_{l=0}^{\infty}{\bf B}_l{\bf r}(t-l,f)
\end{equation}
where ${\bf A}_l \in \mathbb{C}^{M \times N}$ and ${\bf B}_l \in \mathbb{C}^{M \times R}$ are the convolutive transfer functions (CTFs) at the $l$th frame step, $t$ is the frame index and $f$ is the frequency bin index. Since the proposed algorithm is frequency-wise, $f$ is omitted in the following for brevity. 

To extract the direct path and early reflections of the near-end sources, the signal model (\ref{eq:signalmodel}) can be approximately transformed into an auto-regressive model~\cite{takeda2012efficient,togami2014simultaneous} as follows:
\begin{equation}\label{eq:armodel}
  {\bf x}(t) = {\bf A}_0 {\bf s}(t) + {\bar{\bf B}} {\bar{\bf r}}(t) + 
  {\bar{\bf C}} {\bar{\bf x}}(t-\Delta)
\end{equation}
where the delay $\Delta$ marks the boundary between early reflections and late reverberation, and
\begin{align}
\nonumber	{\bar{\bf B}} &= [{\bf B}_0, {\bf B}_1,...,{\bf B}_{L_1-1}], \\
\nonumber	{\bar {\bf r}(t)} &= [{\bf r}(t),{\bf r}(t-1),...,{\bf r}(t-L_1+1)]^T, \\
\nonumber	{\bar{\bf C}} &= [{\bf C}_0, {\bf C}_1,...,{\bf C}_{L_2-1}], \\
	{\bar {\bf x}(t-\Delta)} &= [{\bf x}(t-\Delta),...,{\bf x}(t-\Delta-L_2+1)]^T,
\end{align}
with $L_1$, $L_2$ the orders of transfer functions. 
The matrix notation of (\ref{eq:armodel}) is given by:
\begin{equation}{\label{eq:mixingmodel}}
\begin{bmatrix}
{\bf x}(t)    \\ 
{\bar{\bf r}}(t) \\
{\bar{\bf x}}(t-\Delta)
\end{bmatrix}
= \begin{bmatrix}
{\bf A}_0 & {\bar{\bf B}} & {\bar{\bf C}} \\
{\bf 0} & {\bf I}_1 & {\bf 0} \\ 
{\bf 0} & {\bf 0} & {\bf I}_2
\end{bmatrix}
\begin{bmatrix}
{\bf s}(t) \\ 
{\bar{\bf r}}(t) \\
{\bar{\bf x}}(t-\Delta)
\end{bmatrix}
\end{equation}
where ${\bf I}_1$ and ${\bf I}_2$ are corresponding proper-sized unit matrices.
The upper triangular block mixing matrix in (\ref{eq:mixingmodel}) is invertible if ${\bf A}_0$, the direct path and early reflections transfer function matrix relating the near-end sources and the sensors, is invertible, which is generally true in determined source separation. Hence we assume $M=N$ in the following, and represent the source separation process as:
\begin{equation}\label{eq:w}
\begin{bmatrix}
{\bf s}(t) \\ 
{\bar{\bf r}}(t) \\
{\bar{\bf x}}(t-\Delta)
\end{bmatrix}
=
\underset{{\bf W}}{\underbrace{ 
\begin{bmatrix}
{\bf D} & {\bf E} & {\bf F} \\
{\bf 0} & {\bf I}_1 & {\bf 0} \\ 
{\bf 0} & {\bf 0} & {\bf I}_2
\end{bmatrix}
}}
\begin{bmatrix}
{\bf x}(t)    \\ 
{\bar{\bf r}}(t) \\
{\bar{\bf x}}(t-\Delta)
\end{bmatrix}
\end{equation}
Further assuming the time-independence of the source signals, as has been discussed in~\cite{yoshioka2012generalization,takeda2009ica,takeda2012efficient}, we can use the condition that $\{ {\bf s}(t), {\bar{\bf r}}(t), {\bar{\bf x}}(t-\Delta) \}$ are mutually independent. The separation matrix $\bf W$ can thus be estimated semi-blindly by minimizing Kullback-Leibler Divergence (KLD)
\begin{equation}\label{eq:kld}
	J({\bf W}) = \int p({\bf s},{\bar{\bf r}},{\bar{\bf x}}) \log \frac{p({\bf s},{\bar{\bf r}},{\bar{\bf x}})}{q({\bf s})q({\bar{\bf r}})q({\bar{\bf x}}))} d{\bf s}d{\bar{\bf r}}d{\bar{\bf x}}
\end{equation}
where $p(\cdot)$ is the joint probability density function (PDF) and $q(\cdot)$ the marginal PDFs of the sources.

\section{The proposed algorithm}

\subsection{Joint source separation}

Minimizing (\ref{eq:kld}) is a non-convex optimization problem, to which the auxiliary-function based techniques can be applied instead of the most standard natural gradient approaches~\cite{ono2010auxiliary,ono2011stable}. 
The following joint source separation algorithm is a straightforward extension of the previous work solely on BSS, yet not investigated before.  
The joint algorithm requires that the mixing sources follow a super-Gaussian or generalized Gaussian PDF, which is a valid assumption for speech sources, and the source PDF is represented as:
\begin{equation}\label{eq:sourceprior}
	p(s) \propto \exp [-(\frac{|s|}{\lambda})^\gamma]
\end{equation}
where $\lambda > 0$ and $0 < \gamma \le 2$ denote, respectively, the scaling and shape parameters~\cite{ono2012auxiliary,taniguchi2019generalized}. $\gamma=1$ corresponds to a Laplacian distribution and smaller value yields a more sparse PDF. 

Based on (\ref{eq:sourceprior}), an auxiliary function $J({\bf W, V})$ is designed as
\begin{equation}
	J({\bf W, V}) = \sum_{m=1}^{M}{\bf w}_m^H{\bf V}_{m}{\bf w}_m - \log |\det {\bf W}|
\end{equation}
such that
\begin{equation}
J(\bf W) = \min_{\bf V} J({\bf W, V})
\end{equation}
$(\cdot)^H$ denotes Hermitian transpose. ${\bf w}_m^H$ is the $m$th row vector of $\bf W$, and the introduced auxiliary variable
\begin{equation}\label{eq:vmbss}
	{\bf V}_m = \mathbb{E}[\beta_m(t) {\bf u}(t){\bf u}^H(t)]
\end{equation}
with $\mathbb{E}$ the expectation operator, ${\bf u}(t) = [{\bf s}(t), {\bar{\bf r}}(t), {\bar{\bf x}}(t-\Delta)]^T$, the source PDF related weighting factor
\begin{equation}
\beta_m(t) = (\sum_f |\hat{s}_m(t)|^2)^{\frac{\gamma-2}{2}},
\end{equation}
and the estimate of separated source 
\begin{equation}
\hat{s}_m(t) = {\bf w}_m^H {\bf u}(t).
\end{equation}

The update rule of the separation matrix is given by:
\begin{align}\label{eq:updateW}
\nonumber	{\bf w}_m &= ({\bf W}{\bf V}_{m})^{-1}{\bf i}_m, \\
	{\bf w}_m &= \frac{{\bf w}_m}{\sqrt{{\bf w}^H_m{\bf V}_{m}{\bf w}_m}}
\end{align}
where ${\bf i}_m$ is a one-hot unit vector. The algorithm is then summarized as updating ${\bf V}_m$ and ${\bf W}$ iteratively.

\subsection{Cascaded solutions}

There has always been concern about the computational complexity of the joint algorithms~\cite{takeda2009ica,nakatani2020computationally,ikeshita2021independent}. 
The calculations in (\ref{eq:updateW}) involve matrix multiplication and matrix inversion of order $\mathcal{O}(L^3)$ with $L=M+L_1R+L_2M$, which can be rather computationally expensive for practical applications. An intuitive approach is to decompose the large separation matrix $\bf W$ into smaller parts that can be solved more efficiently. 

An equivalent form of $\bf W$ is given by:
\begin{equation}\label{eq:draec-bss}
{\bf W}_1
= 
\underset{{\bf W}_\text{BSS}}{\underbrace{
\begin{bmatrix}
{\bf D} & {\bf 0} & {\bf 0} \\
{\bf 0} & {\bf I}_1 & {\bf 0} \\ 
{\bf 0} & {\bf 0} & {\bf I}_2
\end{bmatrix}
}}
\underset{{\bf W}_\text{DRAEC}}{\underbrace{
\begin{bmatrix}
{\bf I}_3 & {\bar{\bf E}} & {\bar{\bf F}} \\
{\bf 0} & {\bf I}_1 & {\bf 0} \\ 
{\bf 0} & {\bf 0} & {\bf I}_2
\end{bmatrix}
}}
\end{equation}
where ${\bf E} = {\bf D}{\bar{\bf E}}$ and ${\bf F} = {\bf D}{\bar{\bf F}}$. (\ref{eq:draec-bss}) can be interpreted as performing AEC and DR jointly, and then performing BSS. The corresponding algorithm is denoted as DRAEC-BSS. Taking one step further, we have
\begin{equation}\label{eq:dr-aec-bss}
{\bf W}_2
= 
\underset{{\bf W}_\text{BSS}}{\underbrace{
\begin{bmatrix}
{\bf D} & {\bf 0} & {\bf 0} \\
{\bf 0} & {\bf I}_1 & {\bf 0} \\ 
{\bf 0} & {\bf 0} & {\bf I}_2
\end{bmatrix}
}}
\underset{{\bf W}_\text{AEC}}{\underbrace{
\begin{bmatrix}
{\bf I}_3 & {\bar{\bf E}} & {\bf 0} \\
{\bf 0} & {\bf I}_1 & {\bf 0} \\ 
{\bf 0} & {\bf 0} & {\bf I}_2
\end{bmatrix}
}}
\underset{{\bf W}_\text{DR}}{\underbrace{
\begin{bmatrix}
{\bf I}_3 & {\bf 0} & {\bar{\bf F}} \\
{\bf 0} & {\bf I}_1 & {\bf 0} \\ 
{\bf 0} & {\bf 0} & {\bf I}_2
\end{bmatrix}
}}
\end{equation}
which is denoted as DR-AEC-BSS, and 
\begin{equation}\label{eq:aec-dr-bss}
{\bf W}_3
= 
\underset{{\bf W}_\text{BSS}}{\underbrace{
\begin{bmatrix}
{\bf D} & {\bf 0} & {\bf 0} \\
{\bf 0} & {\bf I}_1 & {\bf 0} \\ 
{\bf 0} & {\bf 0} & {\bf I}_2
\end{bmatrix}
}}
\underset{{\bf W}_\text{DR}}{\underbrace{
\begin{bmatrix}
{\bf I}_3 & {\bf 0} & {\bar{\bf F}} \\
{\bf 0} & {\bf I}_1 & {\bf 0} \\ 
{\bf 0} & {\bf 0} & {\bf I}_2
\end{bmatrix}
}}
\underset{{\bf W}_\text{AEC}}{\underbrace{
\begin{bmatrix}
{\bf I}_3 & {\bar{\bf E}} & {\bf 0} \\
{\bf 0} & {\bf I}_1 & {\bf 0} \\ 
{\bf 0} & {\bf 0} & {\bf I}_2
\end{bmatrix}
}}
\end{equation}
which is denoted as AEC-DR-BSS. 

The above matrix decomposition transforms the separation process in (\ref{eq:w}) into sub-processes that naturally relate to AEC, DR and BSS, which can be solved sequentially instead of jointly. Note that the solving order of BSS is not put first, because it would result in a under-determined source separation sub-problem.

\subsection{Sequential update techniques}

Focusing on the separation matrix (\ref{eq:aec-dr-bss}) in the AEC-DR-BSS algorithm, we first take ${\bf W}_\text{AEC}$ into (\ref{eq:w}) and there is
\begin{equation}\label{eq:semi-aec}
\begin{bmatrix}
{\bf y}(t)    \\ 
{\bar{\bf r}}(t)
\end{bmatrix}
= \begin{bmatrix}
{\bf I}_3       & {\bar{\bf E}}\\ 
{\bf 0} & {\bf I}_1
\end{bmatrix}
\begin{bmatrix}
{\bf x}(t)    \\ 
{\bar{\bf r}}(t)
\end{bmatrix}
\end{equation}
where ${\bf y}(t)$ denotes the reverberant near-end sources, uncontaminated by echo. (\ref{eq:semi-aec}) defines a semi-blind source separation problem, the solution to which has already been provided in our previous work~\cite{wang2020semi,wang2021weighted}. The matrix coefficients are directly given here by:
\begin{align}\label{eq:waec}
{\bar{\bf E}} &= -{\bf Q}_{\text{AEC}}{\bf V}_{\text{AEC}}^{-1}
\end{align}
where
\begin{align}
\nonumber 
{\bf Q}_{\text{AEC}} &= \mathbb{E}[\beta_{\text{AEC}}(t) {\bf x}(t) {\bar{\bf r}}^H(t) ], \\
{\bf V}_{\text{AEC}} &= \mathbb{E}[\beta_{\text{AEC}}(t) {\bar{\bf r}}(t){\bar{\bf r}}^H(t)],
\end{align}
with 
\begin{equation}
   \beta_{\text{AEC}}(t) = |\hat{\bf y}(t)|^{\gamma-2},
\end{equation}
and the estimate of echo canceled source
\begin{equation}
   \hat{\bf y}(t) = {\bf x}(t) + {\bar{\bf E}} {\bar{\bf r}}(t).
\end{equation}

Similarly, there is 
\begin{equation}\label{eq:semi-dr}
\begin{bmatrix}
{\bf z}(t)    \\ 
{\bar{\bf y}}(t-\Delta)
\end{bmatrix}
= \begin{bmatrix}
{\bf I}_3       & {\bar{\bf F}}\\ 
{\bf 0} & {\bf I}_2
\end{bmatrix}
\begin{bmatrix}
{\bf y}(t)    \\ 
{\bar{\bf y}}(t - \Delta)
\end{bmatrix}
\end{equation}
where ${\bar {\bf y}(t-\Delta)} = [{\bf y}(t-\Delta),...,{\bf y}(t-\Delta-L_2+1)]^T$ and ${\bf z}(t)$ denotes the un-reverberant near-end sources. The matrix coefficients are given by:
\begin{align}\label{eq:wdr}
{\bar{\bf F}} &= -{\bf Q}_{\text{DR}}{\bf V}_{\text{DR}}^{-1}
\end{align}
where
\begin{align}
\nonumber 
{\bf Q}_{\text{DR}} &= \mathbb{E}[\beta_{\text{DR}}(t) {\bf y}(t) {\bar{\bf y}}^H(t - \Delta) ], \\
{\bf V}_{\text{DR}} &= \mathbb{E}[\beta_{\text{DR}}(t) {\bar{\bf y}}(t - \Delta) {\bar{\bf y}}^H(t - \Delta)],
\end{align}
with
\begin{equation}
\beta_{\text{DR}}(t) = |\hat{\bf z}(t)|^{\gamma-2},
\end{equation}
and the estimate of the dereverberated source
\begin{equation}
\hat{\bf z}(t) = {\hat{\bf y}}(t) + {\bar{\bf F}} {\bar{\bf y}}(t - \Delta).
\end{equation}

Lastly, the demixing coefficients of $\bf D$ are obtained by applying Aux-IVA to the following problem
\begin{equation}\label{eq:bss}
	{\bf s}(t) = {\bf D} {\bf z}(t),
\end{equation}
and there is the estimate of the desired sources.

Now the DRAEC-BSS and DR-AEC-BSS algorithms can be derived likewise. Note that when solving $\bar{\bf E}$ using (\ref{eq:semi-aec}),  $\bar{\bf F}$ using (\ref{eq:semi-dr}), and ${\bf D}$ using (\ref{eq:bss}), the previous mutual independence assumption of the acoustic echo, late reverberation and clean sources is relaxed to pair-wise independence.

Given the above description, our online implementation of the algorithms involves recursive estimate of the auto-correlation matrix $\bf V$, the cross-correlation matrix $\bf Q$ and the weighting factor $\beta$, using a smoothing coefficient $\alpha$ of 0.999. For the sake of clarity, the source code is available at https://github.com/nay0648/unified2021

\subsection{Complexity analysis}

The cascaded solutions clearly reduce the overall computational cost than the naive joint source separation (Joint-SS) algorithm. A comparison of the order of complexity of the proposed algorithms is shown in Table~\ref{tab:complexity}.

\begin{table}[th]
  \caption{The order of computational complexity of the proposed algorithms.}
  \label{tab:complexity}
  \centering
  \begin{tabular}{ll}
    \toprule
    \text{Algorithm} & \text{Complexity}\\ 
    \midrule
    \text{Joint-SS} & $\mathcal{O}(2ML^3) $  \\ \midrule
    \text{DRAEC-BSS} & $\mathcal{O}(L(L_1R+L_2M)^2+M^3)$  \\ \midrule
    \multirow{2}{*}{\text{DR-AEC-BSS}}  & $\mathcal{O}((M+L_2M)(L_2M)^2+$  \\
                     & $\quad (M+L_1R)(L_1R)^2+M^3)$  \\ \midrule
    \multirow{2}{*}{\text{AEC-DR-BSS}} & $\mathcal{O}((M+L_1R)(L_1R)^2+$  \\
                     & $\quad (M+L_2M)(L_2M)^2+M^3)$  \\
    \bottomrule
  \end{tabular}
\end{table}

\section{Experiments}

\subsection{Setup}

We consider a scenario where one user interacts with a smart speaker in living room environments. The room size is randomly sampled with length in [4.0, 8.0] meters, width in [3.0, 6.0] meters and height in [2.5, 4.0] meters. A microphone array of $M=2$ microphones spacing at 10 cm is placed in the room while keeping a minimum distance of 50 cm to the walls. The $R=1$ loudspeaker playing echo is put 15 cm under the sensor array.
The user and one interfering source ($N=2$) are randomly positioned in the room. Corresponding room impulse responses are generated using the Image method~\cite{allen1979image}.

\begin{figure}[h]
	\centering
	\centerline{\includegraphics[width=0.9\columnwidth]{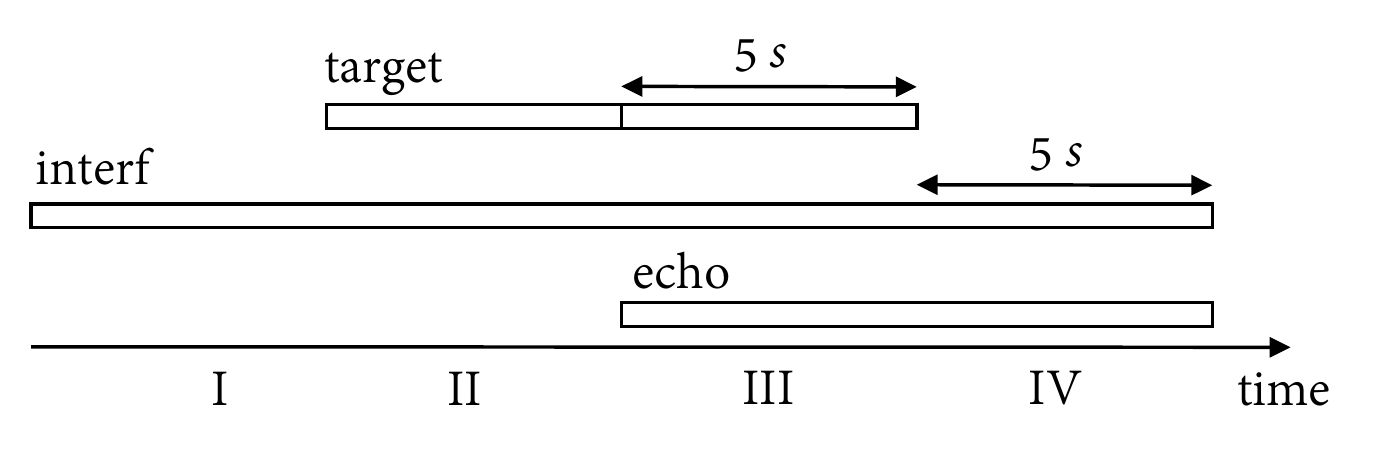}}
	\caption{Source overlapping in the test utterances.}
	\label{fig1}
\end{figure}

The test utterances are prepared following the setup in~\cite{carbajal2020joint}. Specifically, each utterance has four 5-s segments, with the user's speech, interference and echo overlapping as depicted in Figure~\ref{fig1}. 
The input signal-to-interference ratio (SIR) is set at 0~dB, and signal-to-echo ratio (SER) is set at $\{0, -10\}$~dB.
The overall quality of the separated user's speech is measured by signal-to-distortion ratio (SDR)~\cite{vincent2006performance,le2019sdr} in segment III. Two non-instructive metrics, namely signal-plus-interference-plus-echo to interference-plus-echo ratio (SIER) and signal-plus-interference to interference ratio (SIIR), are introduced to measure the non-target reduction performance. 
SIER is roughly estimated as the ratio of signal energy in segment III to that in segment IV. SIIR is estimated as the ratio of signal energy in segment II to that in segment I.
The dereverberation performance is not separately evaluated. Instead, the experiments are repeated under different reverberation time (RT60) of 0.3~s, 0.5~s and 0.8~s.
When calculating the metrics, the direct path and early reflections (50 ms) of the user's speech in the first channel is used as the desired target.

Two classical methods, namely the NLMS algorithm~\cite{valin2007adjusting} for AEC\footnote{https://github.com/wavesaudio/Speex-AEC-matlab} and the WPE algorithm~\cite{yoshioka2010blind} for DR\footnote{http://www.kecl.ntt.co.jp/icl/signal/wpe/index.html} are included using their publicly available implementations and cascaded with the BSS algorithm in Section 3.3 for benchmarking. They are denoted as NLMS-WPE-BSS and WPE-NLMS-BSS.
The test utterances are sampled at 16 kHz. STFT frame size is 1024 and frame shift is 512. The AEC filter tap is set $L_1=5$. The DR filter tap is set $L_2=5$ with a frame delay $\Delta=2$. A sparse source PDF with shape parameter of $\lambda=0.2$ is adopted.

\subsection{Results and analysis}

The SDR, SIER and SIIR improvements with reference to the input mixture are respectively shown in Table~\ref{tab:2}, Table~\ref{tab:3} and Table~\ref{tab:4}. The scores are averaged on 20 independent experiments. 

\begin{table}[th]
	\caption{SDR (dB) improvements with reference to the input mixture in different reverberation time.}
	\label{tab:2}
	\centering
	\begin{tabular}{lcccccc}
		\toprule
		\multirow{2}{*}{\text{Algorithm}} 
		& \multicolumn{3}{c}{SER=0 dB} & \multicolumn{3}{c}{SER=-10 dB} \\
		& 0.3s & 0.6s & 0.8s & 0.3s & 0.6s & 0.8s  \\ \midrule
		\scriptsize{\text{WPE-NLMS-BSS}} 
		& 8.56& 6.75& 5.45& 13.74& 11.62& 10.29  \\ \midrule
		\scriptsize{\text{NLMS-WPE-BSS}} 
		& 8.76& 6.63& 5.51& 14.73& 12.24& 10.79  \\ \midrule
		\scriptsize{\text{Joint-SS}} 
		& 8.76& 6.84& 5.64& 11.86& 9.66& 8.13  \\ \midrule
		\scriptsize{\text{DRAEC-BSS}} 
		& 9.69& 7.50& 6.05& 16.82& 13.54& 12.41  \\ \midrule
		\scriptsize{\text{DR-AEC-BSS}}  
		& 9.51& 7.32& 5.87& 16.05& 12.74& 11.73  \\ \midrule
		\scriptsize{\text{AEC-DR-BSS}} 
		& 9.63& 7.43& 5.97& 16.76& 13.40& 12.32  \\
		\bottomrule
	\end{tabular}
\end{table}

\begin{table}[th]
	\caption{SIER (dB) improvements with reference to the input mixture in different reverberation time.}
	\label{tab:3}
	\centering
	\begin{tabular}{lcccccc}
		\toprule
		\multirow{2}{*}{\text{Algorithm}} 
		& \multicolumn{3}{c}{SER= 0 dB} & \multicolumn{3}{c}{SER= -10 dB} \\
		& 0.3s & 0.6s & 0.8s & 0.3s & 0.6s & 0.8s  \\ \midrule
		\scriptsize{\text{WPE-NLMS-BSS}} 
		& 9.34& 7.25& 5.50& 7.29& 5.42& 5.04  \\ \midrule
		\scriptsize{\text{NLMS-WPE-BSS}} 
		& 9.94& 7.64& 5.82& 8.57& 6.62& 5.24  \\ \midrule
		\scriptsize{\text{Joint-SS}} 
		& 10.35& 8.16& 6.58& 7.29& 5.55& 4.56  \\ \midrule
		\scriptsize{\text{DRAEC-BSS}} 
		& 11.12& 9.25& 7.15& 12.09& 8.93& 7.82  \\ \midrule
		\scriptsize{\text{DR-AEC-BSS}}  
		& 10.71& 8.71& 6.67& 10.40& 6.87& 6.07  \\ \midrule
		\scriptsize{\text{AEC-DR-BSS}} 
		& 10.95& 9.04& 6.82& 11.90& 8.35& 7.21  \\
		\bottomrule
	\end{tabular}
\end{table}

There are clear drops in performance as the reverberation time is larger, where longer filter taps are required for the algorithms. The overall higher scores in SER=-10 dB compared with that in SER= 0dB are due to the baseline scores of the input mixtures, for example, the averaged input SDR is -12.15 dB versus -4.61 dB with RT60=0.3 s.

\begin{table}[th]
	\caption{SIIR (dB) improvements with reference to the input mixture in different reverberation time.}
	\label{tab:4}
	\centering
	\begin{tabular}{lcccccc}
		\toprule
		\multirow{2}{*}{\text{Algorithm}} 
		& \multicolumn{3}{c}{SER= 0dB} & \multicolumn{3}{c}{SER= -10dB} \\
		& 0.3s & 0.6s & 0.8s & 0.3s & 0.6s & 0.8s  \\ \midrule
		\scriptsize{\text{WPE-NLMS-BSS}} 
		& 8.80& 6.35& 5.43& 8.08& 5.72& 4.67  \\ \midrule
		\scriptsize{\text{NLMS-WPE-BSS}} 
		& 8.94& 6.56& 5.70& 8.25& 6.00& 4.97  \\ \midrule
		\scriptsize{\text{Joint-SS}} 
		& 8.27& 6.52& 5.78& 7.49& 5.75& 4.66  \\ \midrule
		\scriptsize{\text{DRAEC-BSS}} 
		& 9.58& 7.89& 6.58& 9.27& 6.97& 6.35  \\ \midrule
		\scriptsize{\text{DR-AEC-BSS}}  
		& 9.33& 7.47& 6.14& 8.62& 6.19& 5.50  \\ \midrule
		\scriptsize{\text{AEC-DR-BSS}} 
		& 9.36& 7.62& 6.24& 9.08& 6.64& 6.05  \\
		\bottomrule
	\end{tabular}
\end{table}

Based on the results here, solving AEC first is better than putting DR first. The conclusion applies both to AEC-DR-BSS and NLMS-WPE-BSS. From the source separation view, the signal independence assumption holds better between echo and near-end sources than that between early reflections and late reverberation. The DRAEC-BSS algorithm performs better than either AEC-DR-BSS or DR-AEC-BSS. 
There could be two reasons. 
The delayed observed signal used in DR could help with more echo reduction. 
And the scaling factor $\beta$, related to spectra of the underlying target source, could be better estimated in DRAEC. 
Joint-SS scores lowest among the proposed algorithms, although it has the highest complexity. This could be due to the poor conditioning of the large covariance matrix as defined in equation~(\ref{eq:vmbss}).

Given the setup used here, the computation cost of DRAEC-BSS is 20\% and AEC-DR-BSS 7\% compared to the Joint-SS baseline.

\section{Conclusion}

This paper considers the tasks of echo cancellation, speech dereverberation and interference suppression from a unified source separation perspective. The Joint-SS algorithm naturally transforms into cascades of the separate AEC, DR and BSS algorithms, and their solving orders impact the final performance. The proposed DRAEC-BSS solution not only reduces largely the computational cost but also shows better capability than the other setups.

\bibliographystyle{IEEEtran}

\bibliography{mybib}

\end{document}